\documentclass[submission, Phys]{SciPost}

\hypersetup{
    colorlinks,
    linkcolor={red!50!black},
    citecolor={blue!50!black},
    urlcolor={blue!80!black}
}

\usepackage[bitstream-charter]{mathdesign}
\usepackage{subcaption}
\usepackage{float}
\usepackage[normalem]{ulem} 
\usepackage{tikz}
\usepackage{graphicx}
\usetikzlibrary{positioning}

\definecolor{olivegreen}{RGB}{128,128,0}

\DeclareSymbolFont{usualmathcal}{OMS}{cmsy}{m}{n}
\DeclareSymbolFontAlphabet{\mathcal}{usualmathcal}

\begin{document}

\begin{center}{\Large \textbf{
Experimental Investigation of a Bipartite Quench \\ in a 1D Bose gas\\
}}\end{center}

\begin{center}
L. Dubois\textsuperscript{1}, G. Themèze\textsuperscript{1}, J. Dubail\textsuperscript{2} and I. Bouchoule\textsuperscript{1},
\end{center}

\begin{center}
{\bf 1} Laboratoire Charles Fabry, Institut d'Optique, CNRS, Université Paris-Saclay
 \\
 {\bf 2} CESQ and ISIS (UMR 7006), University of Strasbourg and CNRS, 67000 Strasbourg, France
\\
${}^\star$ {\small \sf lea.dubois@universite-paris-saclay.fr}
\end{center}

\begin{center}
\today
\end{center}


\section*{Abstract}
{\bf
Long wavelength dynamics of 1D Bose gases with repulsive contact interactions can be captured by Generalized HydroDynamics (GHD) which predicts the evolution of the local rapidity distribution. The latter corresponds to the momentum distribution of quasiparticles, which have infinite lifetime owing to the integrability of the 
system.
 Here we experimentally investigate the dynamics for an  initial situation that is the junction of two 
semi-infinite systems in different stationary states, a protocol referred to as `bipartite quench' protocol.
More precisely  we realise the particular case where one half 
of the system is the vacuum state. 
 We show that the evolution of the boundary density profile exhibits ballistic dynamics obeying the Euler hydrodynamic scaling. The boundary profiles are similar to the ones predicted with zero-temperature GHD in the quasi-BEC regime, with deviations due to non-zero entropy effects. 
 We show that  this protocol, provided the boundary profile is measured with infinite precision, permits to reconstruct the rapidity distribution of the initial state. 
 For our data, we extract the initial rapidity distribution by fitting the boundary profile and we use 
 a 3-parameter ansatz that goes beyond the thermal assumption. 
Finally, we investigate the local rapidity distribution inside the boundary profile, which, according to GHD, presents, on one side, features of  zero-entropy states. The measured distribution shows the asymmetry predicted by GHD, although unelucidated deviations  remain. 
}

\vspace{10pt}
\noindent\rule{\textwidth}{1pt}
\tableofcontents\thispagestyle{fancy}
\noindent\rule{\textwidth}{1pt}
\vspace{10pt}

\section{Introduction} 

\label{sec:intro}

Gaining insight on the out-of-equilibrium dynamics of many-body quantum systems is tremendously difficult and is the goal of an active research field.  
One particular class of systems where important progress has been made over the past decade is the class of integrable one-dimensional systems. 
Owing to their infinite number of local conserved charges, the description of the local properties of stationary states that arise after relaxation requires a whole function, the rapidity distribution~\cite{yang1969thermodynamics,zamolodchikov1990thermodynamic,rigol2007relaxation,mossel2012generalized,caux2012constructing,fagotti2013reduced,ilievski2016string}.
This function can be viewed as the velocity distribution of the infinite-lifetime quasi-particles in the system.   Its large-scale effective dynamics is described by  generalized hydrodynamics (GHD) \cite{bertini_transport_2016,castro-alvaredo_emergent_2016} (for recent reviews, see e.g.~\cite{bastianello2022introduction,doyon2025generalized}), which assumes local relaxation to a local stationary state.
As with any hydrodynamic theory, the most paradigmatic situation that can be handled by GHD is the `Riemann problem' \cite{riemann1860fortpflanzung}, also dubbed `bipartite quench' more recently~\cite{de_nardis_edge_2018,horvath2019hydrodynamics,alba2019entanglement,rylands2023transport,gamayun2023landauer,horvath2024full}, or `domain-wall quench' or `domain-wall protocol'~\cite{yuan2007domain,antal1999transport,hauschild2016domain,collura2018analytic,misguich2019domain,scopa2022exact,collura2020domain,scopa2023scaling,mcroberts2024domain}. In this `bipartite quench protocol', the microscopic dynamics is governed by a translation-invariant Hamiltonian but the initial state is the junction of two semi-infinite homogeneous systems each prepared in a different stationary state of the Hamiltonian. The GHD theory predicts that, at times long enough such that diffusion effects become negligible~\cite{de_nardis_diffusion_2019} and Euler-scale hydrodynamics is valid, the time evolution is ballistic. 
An interesting feature of this protocol is that the local state, within the boundary, is expected to present features characteristic of zero-temperature systems. Thus, this protocol could be used to reveal power-law singularities of correlation functions characteristic of a zero-temperature Luttinger liquid~\cite{de_nardis_edge_2018}, provided a local probe is used. 

In this paper, we experimentally realize an instance of the bipartite quench protocol using an ultra-cold atomic Bose gas, well described by the Lieb-Liniger model of one-dimensional Bosons with contact repulsive interactions\cite{lieb_exact_1963,bouchoule_generalized_2022}, which is an integrable model. 
In our experiment, the bipartition consists of the junction of a gas in a stationary state on one side, and the vacuum on the other side. This is in contrast with another  bipartite quench protocol realized very recently in the Lieb-Liniger gas in Ref.~\cite{schuttelkopf2024characterising}. There, a completely different method is used, which allows to create an initial state with different non-vanishing densities on the left and on the right. In our work, the initial state is prepared by producing a homogeneous atomic cloud and by removing suddenly its left part. For different evolution times, we record the density profile of the boundary between the two regions, dubbed the boundary profile.  We find that the boundary profile exhibits a ballistic behavior, as expected from the predictions of GHD theory at the Euler scale. 

The boundary profile, for clouds prepared with deep evaporative cooling, is in fair agreement with GHD predictions assuming the semi-infinite gas is in its ground state, although deviations are present. From the boundary profile, we show that it is in principle possible to reconstruct the rapidity distribution characterizing the initial gas. This protocol can thus be used as a generalized thermometry. 
However, the reconstruction method suffers from a high sensitivity to experimental noise in the tail of the boundary profile, which prevents us from reconstructing faithfully the initial rapidity distribution. Instead, we use an ansatz parametrized by a few parameters to extract the rapidity distributions of the initial gas from a fit to the boundary profile. 

Finally, we use a newly developed technique \cite{dubois_probing_2024} to probe the local rapidity distribution within the boundary. The latter is expected to be highly asymmetric for an initial state whose rapidity distribution is substantially broader and 
smoother than that of the ground state: while one of its borders
reflects the broad character of the initial rapidity distribution, the other border
presents the sharp feature expected for the ground state. Our experimental data show such an asymmetric behavior, although the sharp border presents an unelucidated tail.

\section{Experimental setup}

We produce an ultra-cold gas of $^{87}$Rb bosonic atoms in the stretched state $|F=2,m_F=2 \rangle$ using an atom chip. In addition to a homogeneous longitudinal magnetic field $B_0 = 3.36 $G, transverse trapping is achieved with three parallel microwires deposited on the chip (shown in blue in Fig.\ref{fig:setup}(a)) which carry AC currents modulated at $400$MHz. This configuration eliminates wire roughness effects and allows independent control over both longitudinal and transverse confinement \cite{PhysRevLett.98.263201}. The atoms are trapped $7\mu$m from the chip surface and $15\mu$m from the wires, enabling strong transverse confinement. The transverse trapping potential is well approximated by a harmonic potential with a frequency of $\omega_{\perp}/ 2 \pi=2.56 $kHz. 
Using radio-frequency evaporative cooling, we produce atomic clouds whose linear density $n_0$ is close to 50 at/$\mu$m.
Further details on the setup can be found in \cite{duboistel-04749900}. 
The effective 1D coupling constant for atoms in the transverse ground state is given by $g = 2 a_{3D} \hbar \omega_{\perp}$ \cite{PhysRevLett.81.938}, where $a_{3D}=5.3$ nm is the $3$D scattering length of $^{87}$Rb~\cite{PhysRevLett.89.283202}. 
The dimensionless Lieb parameter $\gamma =  mg / (\hbar^2 n_0)$
lies in the range $ [0.4,0.7] \times 10^{-2}$ for the data presented in this paper, corresponding to an interaction energy $gn_0$ in $ [0.4\hbar\omega_\perp,0.6\hbar\omega_\perp] $. 
From our experience of 
radio-frequency cooling in our  
set-up~\cite{dubois_probing_2024,johnson_long-lived_2017}, we expect
the cloud to lie in the quasi-condensate regime\cite{PhysRevLett.91.040403}, with a  temperature 
ranging from about $gn_0$ to  a few times $gn_0$.

The longitudinal magnetic trap is produced by DC currents
running through four wires positioned on either side of the three microwires, as shown in the Fig.\ref{fig:setup}(a). Since these wires are placed far from the atomic cloud, the longitudinal potential can be expressed as a polynomial series expansion $V(x)= \sum_{i} a_i x^i$. 
The fourth first coefficients $a_i$ are tuned by adjusting the currents in the four wires that generate the longitudinal trapping potential. By carefully selecting these currents, it is possible to set $a_1$, $a_2$ and $a_3$ to zero such that 
the leading term is  the quartic term $V(x)=a_4 x^4$.
Such a potential permits to achieve a quasi-homogeneous atomic density over a relatively large region, an important feature to study the bipartite quench protocol which assumes 
a semi-infinite system.
An example of linear density profile for an atomic cloud placed in such a potential is represented in gray in Fig.\ref{fig:setup}(b). The linear density $n_0$ remains constant to within $10 \%$ around the peak density over a range of approximately $250 \mu$m.

To produce the initial bipartition, we use the selection method introduced in \cite{PhysRevLett.133.113402}. We illuminate the left border of the atomic cloud, initially in a global stationary state in a quartic trap, with a pushing beam that is nearly resonant with the $F=2 \to F' = 3$ transition of the $D2$ line and which propagates perpendicularly to $x$. Atoms shined by this pushing beam are subjected to radiation pressure : after being illuminated for $30 \mu$s corresponding to $\sim 15$ absorption/reemission cycles, atoms
have enough energy to leave the trap. To illuminate only a border of the gas, the beam is shaped using a digital micromirror device (DMD). Further details on this spatial selection method are available in \cite{PhysRevLett.133.113402}. This protocol produces a sharp boundary between a zero density system and a quasi-homogeneous gas due to the fact that the atoms are initially placed in a quartic trap.  The sharpness of the boundary is mainly limited by the imaging resolution, which is in the micrometer range. The reabsorption of scattered photons by the atoms which are not shined could also limited the boundary sharpness. This effect is mitigated by detuning the pushing beam by $15$MHz from the D$2$ transition. An example of the density profile of a gas initially in a global stationary state in a quartic trap, after applying this spatial selection tool, is shown in yellow in Fig.\ref{fig:setup}(b).

The longitudinal confinement is then removed while maintaining the transverse confinement. The initial sharp boundary broadens in time 
and this  dynamics is monitored by recording longitudinal density profiles $n(x,t)$ after different evolution time $t$. 
\begin{figure}[!htb]
    \centering
    \includegraphics[width=0.90\linewidth]{Chip_selection_V4.pdf}
    \caption{(a) Schematic drawing of the atom chip. The $3$ blue wires produce the transverse trapping, the $4$ other wires produce the longitudinal trapping. The red oval ball represents the atomic cloud, trapped 12 microns above the wires $-$ (b) Linear density profiles extracted from absorption images. The gray curve is the linear density profile of gas confined within a quartic potential. The atomic cloud is then illuminated during $30 \mu$s by a near resonant light beam, shaped using a DMD. The resulting density profile after a time of flight of $1$ms is depicted in yellow. }
    \label{fig:setup}
\end{figure}

\section{GHD predictions}\label{sec.GHDpredictions}
\label{sec:ghd}

The above experimental setup is well described theoretically by the  GHD approach~\cite{schemmer2019generalized,malvania_generalized_2021} which works as follows. Under time evolution, the initial sharp boundary of the cloud gets smoother, and the time
derivatives of local quantities decrease. 
After some time, upon coarse-graining, one expects that the gas can locally be described by stationary states.
Stationary states of the Lieb-Liniger model are entirely characterized 
by their rapidity distribution $\rho(\theta)$.
Equivalently, they can be characterized by a function $\nu(\theta)$ dubbed `occupation factor' which takes values between 0 and 1, and which is related to  
the rapidity distribution $\rho(\theta)$ by
\begin{equation}
\nu (\theta)=\frac{\rho(\theta)}{\rho_s(\theta)} \mbox{, \qquad where \qquad } \rho_s(\theta)=
\frac{m}{2\pi \hbar} + \int \frac{d\theta'}{2\pi} \Delta(\theta-\theta') \rho(\theta') ,  \end{equation}
and $\Delta(\Theta)=2g/(g^2/\hbar+\hbar\Theta^2)$ is the `scattering shift' in the Lieb-Liniger model. The functions $\nu$ and $\rho$ are in one-to-one correspondence and in the following we use alternately $\rho$ or $\nu$. [For an introduction to this formalism, we refer to the lecture notes \cite{doyon2020lecture} or to Section 1 of the review article \cite{bouchoule_generalized_2022}.]

Since we assume local stationarity, the system as a whole is described by a time- and position-dependent rapidity distribution $\rho(x,t,\theta)$, or equivalently by the 
time- and position-dependent occupation factor $\nu(x,t,\theta)$. The latter leads to simpler calculations, while the former is particularly useful to extract the linear density, which reads
\begin{equation}
    \label{eq:lineardensity}
n(x,t)=\int d\theta \, \rho(x,t,\theta)  .
\end{equation}

\begin{figure}[!htb]
    \centering
    \includegraphics[width=0.6\linewidth]{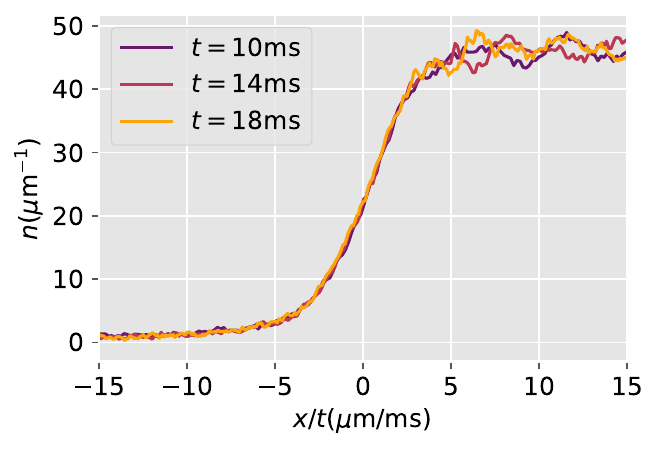}
    \caption{Test of the ballistic scaling. Boundary density profiles obtained for different evolution times $t$ and represented as a function of $x / t$.
    The profiles overlap remarkably well, showing that the Euler scale is reached within this time interval. 
    The longitudinal dynamics after $t = 18$ms cannot be probed due to the fact that our initial semi-homogeneous gas has a finite size. For shorter deformation times, experimental  boundary profiles are smoother than the Euler-scale GHD predictions, which might be due to the failure of Euler scale, and/or to the fact that the cut at $t=0$ is not infinitely sharp. }
    \label{fig:euler}
\end{figure}

The GHD equations \cite{bertini_transport_2016,castro-alvaredo_emergent_2016} predict the time evolution of $\rho(x,t,\theta)$, or equivalently of $\nu (x,t,\theta)$. When written in terms of the occupation factor $\nu (x,t,\theta)$, the GHD equations take the form of a convective equation 
\begin{subequations}
\label{eq:GHD}
\begin{equation}
\frac{\partial\nu}{\partial t} + v^{\rm{eff}}_{[\nu]}\frac{\partial  \nu }{\partial x} = 0 ,
\end{equation}
and a second relation that fixes the effective velocity $v^{\rm{eff}}_{[\nu]}$ as a functional of the local rapidity distribution,
\begin{equation}
v_{[\nu]}^{\rm{eff}}(\theta) = \theta -\int  \Delta(\theta-\theta') \left (  v^{\rm{eff}}_{[\nu]}(\theta) - v^{\rm{eff}}_{[\nu]}(\theta') \right ) \rho(\theta') d\theta' .
\end{equation}
\end{subequations}
More precisely, Eq.~(\ref{eq:GHD}a) is the `Euler-scale' form of GHD, a diffusionless equation that is valid at the large scales. Diffusive corrections that enter in the form of a Navier-Stokes-type term~\cite{de2018hydrodynamic,de2019diffusion,bastianello2020thermalization,de2022correlation}, or even dispersive corrections~\cite{de2023hydrodynamic}, have also been studied theoretically. However, they are subleading and so far they have not been observed experimentally. We will see below that our experimental data obey the scaling collapse expected at the Euler scale (Fig.~\ref{fig:euler}), so these effects seem to be negligible in our situation, at least for the analysis of the boundary profiles. This is also compatible with a recent theoretical study in the weakly interacting regime that has concluded that diffusive effects should be very small~\cite{moller2024identifying}. Therefore, in this paper we ignore the possibility of subleading diffusive effects (as well as all higher-order effects) in our modeling, and we stick to the Euler-scale GHD equation above.

For an initial bipartition whose discontinuity is located at $x=0$, the solution of \eqref{eq:GHD} depends on $x$ and $t$ only through the ratio $\zeta=x/t$~\cite{bertini_transport_2016,castro-alvaredo_emergent_2016}. Thus all local properties of the gas should depend only on the ratio $\zeta$. In particular the density profile should be a function of $\zeta$ only. We have checked that our experimental data verify that ballistic scaling, see Fig.~\ref{fig:euler}.

\begin{figure}[hbt]
    \centering
    \begin{tikzpicture}
        \draw (0,0) node {\includegraphics[width=0.7\textwidth]{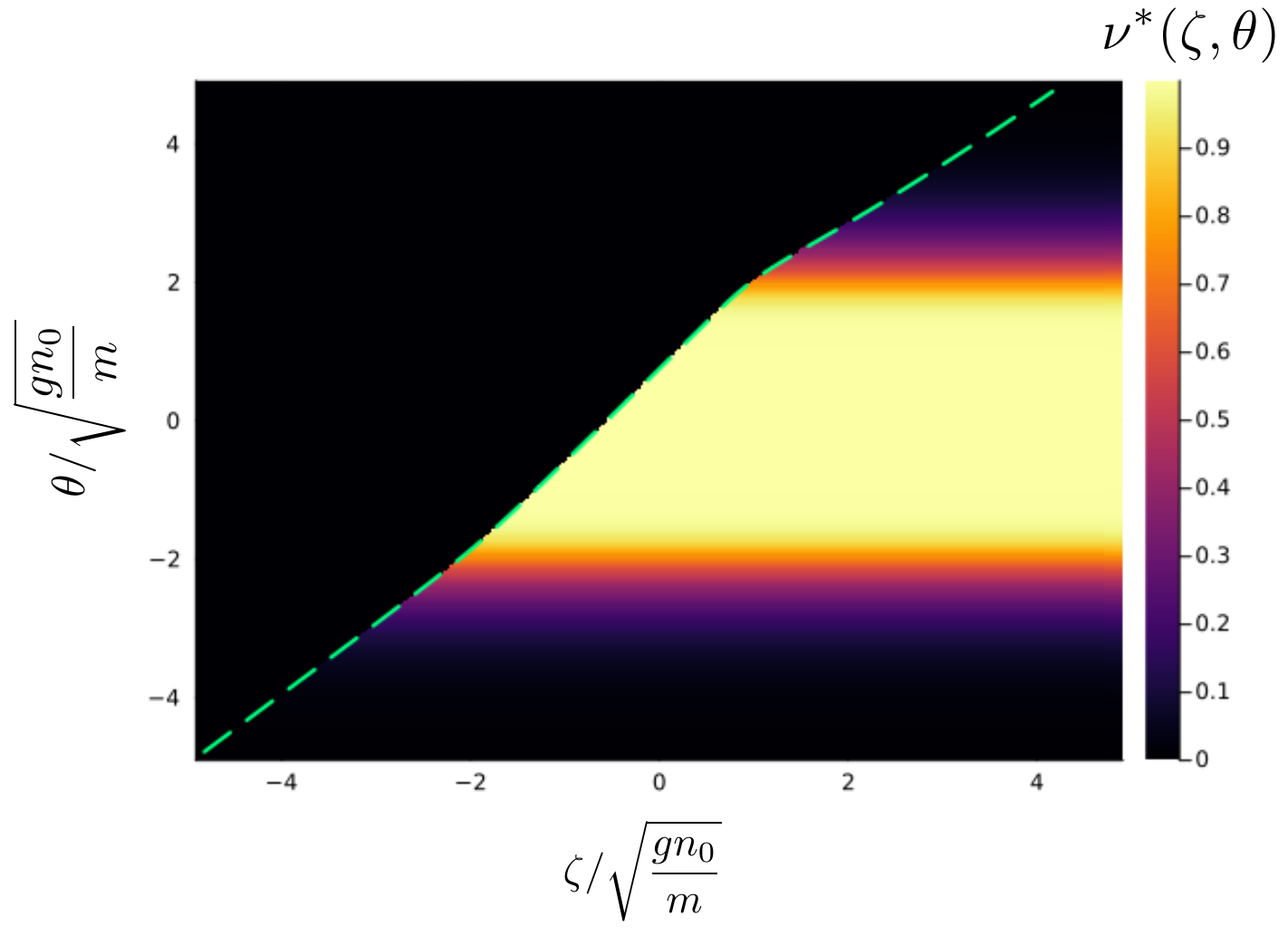}};
        \draw[red,dashed, thick] (0,3) -- (0,-2.5);
        \draw[red,dashed, thick] (0.4,3) -- (0.4,-2.5);
        \draw[red] (0.2,4.2) node {slice};
        \draw[red] (0.2,3.8) node {\small (Section 6)};
        \draw[red,thick, ->] (0.2,3.6) -- (0.2,3.3);
    \end{tikzpicture}
    \caption{Occupation ratio $\nu^* (\zeta,\theta)$ solving the equation (\ref{eq:nuetoile}) for an initial occupation ratio $\nu_0 (\theta)$ in the right half-system corresponding to thermal equilibrium at temperature $T$. The dashed green line is the curve $\theta^*(\zeta)$, {\it i.e.} it is the set of points $(\zeta,\theta)$ such that $v^{\rm eff}_{[\nu^* (\zeta,.)]} (\theta) = \zeta$. [Parameters: $\gamma_0=mg/(n_0\hbar^2)=0.005$, $k_B T \hbar^2/(mg^2) = 365$, close to the experimental parameters of the data set of Fig.~\ref{fig:fitted_border}.] The two vertical red dashed lines show a typical `slice' of the boundary profile, which we study in detail in Section~\ref{sec:local}. In that slice, the occupation factor is highly asymmetric: it varies smoothly with $\theta$ for negative values of $\theta$, while it behaves as a step function for $\theta$ close to $\theta^*$.}
    \label{fig:nu_star}
\end{figure}

Let us now elaborate on the solution of Eq.~\eqref{eq:GHD}. According to the above 
scaling, it is of the form
\begin{equation}
\label{eq:nuvsnuetoile}
    \nu(x,t,\theta)=\nu^*( x/t,\theta)
\end{equation} 
for some function $\nu^*( x/t,\theta)$. For the situation considered in this paper with, initially, a vacuum state for $x<0$ and a state of occupation factor $\nu_0 (\theta)$ for $x>0$,  the solution $\nu^*(\zeta,\theta)$ is parameterized by an edge rapidity $\theta^*$ according to~\cite{bertini_transport_2016,castro-alvaredo_emergent_2016}
\begin{equation}
\label{eq:nuetoile}
    \nu^*(\zeta,\theta)=\left \{ \begin{array}{ccc} 
    \nu_0(\theta) &\mbox{ if }& \theta < \theta^*\\
    0 & \mbox{ if } & \theta > \theta^*\\
    \end{array} \right . \mbox{ where  \quad }  v^{\rm{eff}}_{[\nu^*(\zeta,.) ]}(\theta^*)=\zeta \, .
\end{equation}
This equation can be solved numerically 
for any given initial distribution $\nu_0(\theta)$, see Fig.~\ref{fig:nu_star} for an example. Together with Eq.~\eqref{eq:nuvsnuetoile}, it entirely describes the system at the Euler scale. Note that, to compute the linear density $n(x,t)$ in order to compare with experimental density profiles,  one uses Eq.~(\ref{eq:lineardensity}).

\section{Solution for a cloud initially in the ground state}

To illustrate the above formalism, let us explore its implications for the special case where the right half-system is initially in the ground 
state. In that case the initial occupation factor $\nu_0 (\theta)$ is a Fermi sea: $\nu_0(\theta)=1$ for $|\theta| < \Delta\theta_0$, and $\nu_0(\theta)=0$ otherwise. The Fermi radius $\Delta\theta_0$ depends on the initial linear density $n_0$ through Eq.~(\ref{eq:lineardensity}) \cite{lieb_exact_1963}.

\begin{figure}[thb]
    \centerline{
    \includegraphics[height=0.3\textwidth]{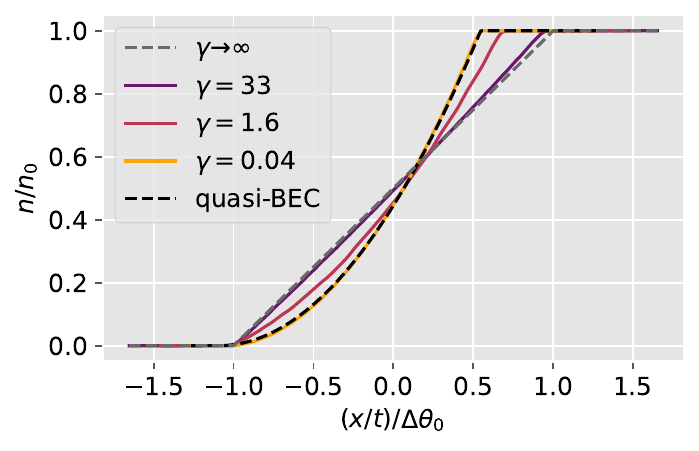}
    \includegraphics[height=0.3\linewidth]{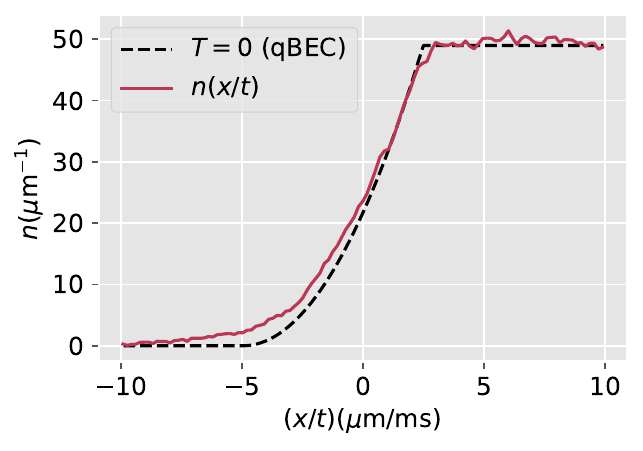}}
    \caption{(a) Boundary profile predictions from GHD for system initially in the ground state as a function of $\gamma$. The velocity  is normalized to the radius of the intial Fermi sea $\Delta \theta_0$. 
    On the negative side the 
    point where $n$ reaches 0 is at $\Delta \theta_0$ whatever $\gamma$.
    On the positive side, the point where $n$ reaches $n_0$ is at 
    the speed of sound $c$.
    The black, resp. grey, dashed line corresponds to the hydrodynamic prediction in the
    quasi-BEC regime (Eq.~\eqref{eq:GPE}), resp. in the  hard-core regime  (Eq.\eqref{eq:HS}). (b) Comparison between experimental data and zero temperature prediction. The latter is given by eq. \eqref{eq:GPE} with very good precision since the interaction parameter of the data is as low as $\gamma = 4.6\times 10^{-3}$. [The experimental curve, recorded for an evolution time $t=10$ ms, belongs to a different data-set than that used in Fig.~\ref{fig:euler}.]}
    \label{fig:boundary_profiles_theory}
\end{figure}

In that case the general features of the function $\nu^*(\zeta,\theta)$ that solves the GHD equation (\ref{eq:GHD})  are as follows ( see Fig.~\ref{fig:boundary_profiles_theory}). It comprises three regions: an `empty region' far on the left with vanishing atom density, a `filled  region' far on the right where the density is equal to the initial density $n_0$, and a `central region' where the atom density interpolates between $0$ and $n_0$. It is easy to see that the left endpoint, where the atom density vanishes, is at $\zeta= -\Delta\theta_0$. 
The right endpoint velocity on the other hand is the sound velocity in the fluid of density $n_0$, given by $c =v^{\rm eff}_{[\nu_0]} (\Delta \theta_0)$.
In the central region $ - \Delta \theta_0 < x/t < c$, the gas is locally in a state that is a Fermi sea shifted by a Galilean boost of velocity $V(x/t)$ for some function $V$. For arbitrary interaction strengths $g$ 
the density profile $n(x/t)$ cannot be computed in closed form, but it is easily computed numerically. Analytical expressions are available in the two asymptotic regimes of strong and weak interactions which correspond to $\gamma\gg 1$ and $\gamma\ll1$ respectively.

In the strong repulsion regime, or hard-core regime, 
$v^{\rm eff}_{[\nu]} (\theta)=\theta$ regardless of the occupation factor $\nu(\theta)$. Then 
Eq.\eqref{eq:nuetoile} is easily solved. 
We can use the fact that, in this regime, a Fermi sea of radius $\Delta\theta$ corresponds to a linear density $n=m\Delta\theta/(\pi\hbar)$ to derive

\begin{equation}
    (\gamma \gg 1) \qquad \qquad  n(x,t) \, = \, 
    \frac{n_0}{2}  \left( 1 +  \frac{xm}{t\pi\hbar n_0} \right)   \qquad {\rm if} \quad   -\pi\hbar n_0/m < x/t  < \pi\hbar n_0/m.
    \label{eq:HS}
\end{equation}

We can easily check that we recover the result expected for a gas of free fermions, as expected from the mapping of the hard-core bosons to fermions, which preserves the density\cite{girardeau_relationship_1960}.

In the weakly interacting regime, or quasi-BEC regime, the effective velocity at the edge of a Fermi sea of radius $\Delta\theta$ is 
$\Delta\theta/2$, in the frame where the Fermi sea is at rest. Then, using the fact that, in this regime, a Fermi sea of radius $\Delta\theta$ corresponds to a linear density $n=m\Delta\theta^2/(4 g)$, we obtain
\begin{equation}
   (\gamma \ll 1) \qquad \quad  n(x,t)= 
    n_0 \left ( \frac{2}{3}+ \frac{1}{3} \frac{x}{ t}\sqrt{\frac{m}{gn_0}} \right )^2 \qquad  {\rm if} \quad  -2\sqrt{gn_0/m}   < x/t < \sqrt{gn_0/m}.
    \label{eq:GPE}
\end{equation}

Here we recover the hydrodynamic predictions derived from the Gross-Pitaevskii equation\cite{el_decay_1995,xu_dispersive_2017}. This is expected, since the Gross-Pitaevskii approach becomes exact in the limit of weak interactions, so it should agree with GHD, because GHD is the correct hydrodynamic equation for all repulsion strengths.

In Fig.~\ref{fig:boundary_profiles_theory}(a), we compare the  GDH solution for 
systems initially in the ground state with the two above asymptotic formulas. We find that the quasi-BEC regime is reached to an excellent approximation already for $\gamma = 0.04$. 

Fig.~\ref{fig:boundary_profiles_theory}(b) compares  the  measured boundary profile $n(\zeta)$ to the GHD prediction assuming that the initial state on the right is the ground state. Here the Lieb parameter is $\gamma = 4.6\times 10^{-3}$, and for such small values the boundary profile is indistinguishable from the quasi-condensate prediction, see Fig.~\ref{fig:boundary_profiles_theory}, so we actually compare the data to Eq.~\eqref{eq:GPE}. 
The agreement between the ground state prediction and experimental data 
is rather good, especially in the high density part. 
The deviations from the parabola observed experimentally are  due to non-zero entropy effects, which are investigated  in the following section.

\section{Retrieving the initial rapidity distribution from the boundary profile}
\label{sec:retreving}

In Section~\ref{sec:ghd} we saw that, for a given initial occupation factor $\nu_0(\theta)$, we can compute the boundary profile $n(\zeta)$ with the Euler-scale GHD equations. Since, in the experiment, we measure the boundary profile, it is natural to ask whether the converse operation is possible: Can we retrieve the occupation factor $\nu_0(\theta)$ from the boundary profile, relying on the Euler-scale GHD equations?

\paragraph{Direct reconstruction.}
We assume that we have a boundary profile $n(v)$ which is monotonically increasing, with $n(\zeta) = 0$ when $\zeta \rightarrow -\infty$, and $n(\zeta) = n_0$ when $\zeta \rightarrow +\infty$. 
A first idea is to reconstruct the function $\nu_0(\theta)$ incrementally, from negative values of $\theta$ to positive ones, by `reading' the boundary profile $n(\zeta)$ from left to right. We can start from some highly negative velocity $\zeta_0$, such that $n(\zeta)$ is extremely small for all $\zeta \leq \zeta_0$ so it can be assumed to (numerically) vanish: $n(\zeta) = 0$ for all $\zeta \leq \zeta_0$. We work with discrete values of the rapidities, with a constant spacing $\delta \theta > 0$, 
\begin{equation}
    \theta_j  =  \zeta_0 + j \delta \theta , \qquad j \in \mathbb{N} ,
\end{equation}
and we reconstruct the corresponding values of the occupation factor $\nu_j$ ($\simeq \nu_0(\theta_j)$) inductively. We initialize the sequence as
\begin{equation}
    \nu_0 = 0 .
\end{equation}
At the $j^{\rm th}$ step, all the occupation factors $\nu_0, \nu_1, \dots, \nu_{j-1}$ are known, and we want to compute $\nu_j$. We fix $\nu_{j}$ by requiring that
\begin{equation}
    n (\zeta_{j} ) \, = \, n_j , 
    \label{eq:recursif}
\end{equation}
where $n(\zeta)$ is the given boundary profile, and $n_j$ and $\zeta_j$ are numerical estimates of the particle density and of the effective velocity respectively, obtained by discretizing the various integrals that enter the definitions of Section~\ref{sec:ghd}:
$$
n_j = \sum_{a=0}^j \frac{\delta \theta}{2\pi} \nu_a  1^{\rm dr}_{j,a} \quad  \left( \simeq \int^{\theta_j}_{-\infty} \frac{d\theta}{2\pi} \nu(\theta) 1^{\rm dr}(\theta) \right)
$$
$$
\zeta_j = \frac{{\rm id}^{\rm dr}_{j,j}}{1^{\rm dr}_{j,j}}  \quad \left( \simeq \frac{{\rm id}^{\rm dr} (\theta_j)}{1^{\rm dr} (\theta_j)} \right) .
$$
Here ${\rm id}(\theta) = \theta$ is the identity function, and the discretized dressed function $f^{\rm dr}_j$, for a function $f$, is the solution of the linear system
$$
f^{\rm dr}_{j,a} = f (\theta_a) + \sum_{b=0}^j \frac{\delta \theta}{2\pi} \Delta(\theta_a - \theta_b) \nu_b  f^{\rm dr}_{j,b} .
$$
This is the discrete analog of the definition of the dressing, which is the solution of the integral equation $ f^{\rm dr}(\theta) = f(\theta) + \int_{-\infty}^{\theta_j} \frac{d\theta'}{2\pi} \Delta (\theta-\theta') \nu(\theta') f^{\rm dr}(\theta') $ [see e.g. Refs.~\cite{doyon2020lecture,bouchoule_generalized_2022} for introductions to this formalism]. The value of $\nu_j$ that fulfills Eq.\eqref{eq:recursif} can be found numerically with a root-finding algorithm; we use the bisection method.

In the limit of small spacing $\delta \theta$, this procedure is expected to converge to a continuous occupation factor $\nu_0(\theta)$. Numerical tests presented in the appendix confirm that this method retrieves the correct $\nu_0(\theta)$.


However, when we try to apply this method to experimental boundary profiles, we face two difficulties. First, from spares  and noisy experimental data points one needs to extract an increasing continuous function $n(\zeta)$. For this, we need to fit the data with some ansatz for the boundary function. Second, we have observed that this method is highly sensitive to the details of the boundary profile $n(\zeta)$, especially to the left tail of $n(\zeta)$ at negative values of $\zeta$. Since the signal-to-noise ratio in our experimental data is poor in this region, the results obtained with this technique are not trustworthy -- see details in the appendix. Thus, we prefer to use an alternative method, which we present now.

\paragraph{Fitting the occupation factor $\nu_0(\theta)$.}
In order to extract the occupation factor distribution $\nu_0 (\theta)$, we fit the experimental boundary profile with the GHD calculations based on Eqs.~(\ref{eq:nuvsnuetoile})-(\ref{eq:nuetoile}). Extracting $\nu_0(\theta)$ exactly would correspond to a fit with infinitely many fitting parameters, which we are not able to do. So we choose an ansatz for $\nu_0(\theta)$, parameterized only by a few fitting parameters.

\begin{figure}[!htb]
    \centerline{
    \includegraphics[width=\linewidth]{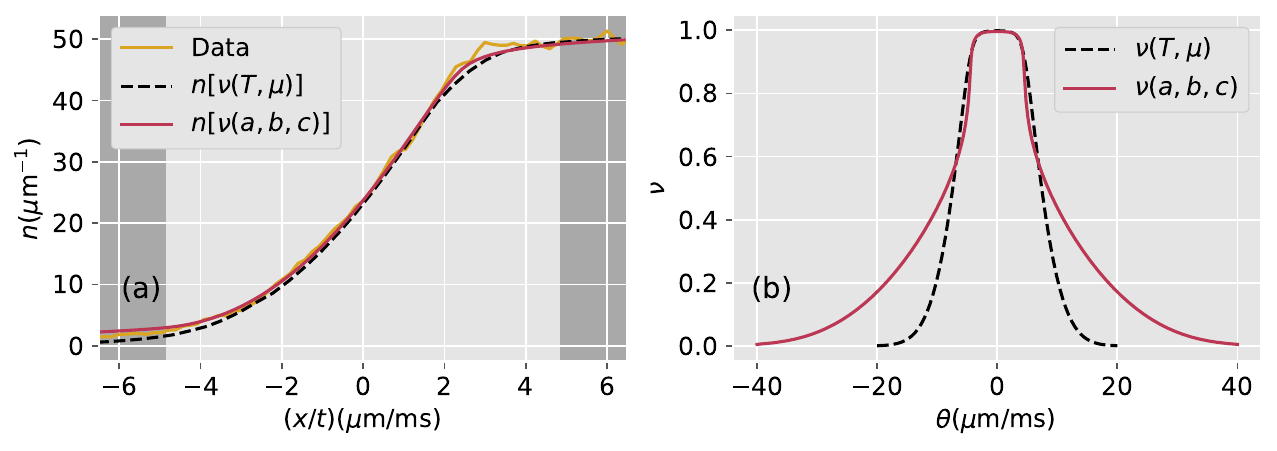}}
    \caption{
    (a) The experimental boundary profile plotted in yellow is compared to fitted profiles using
    for $\nu_0(\theta)$ either  a thermal ansatz, {\it i.e.} the solution of Eqs.~\eqref{eq:fonctions} and  \eqref{eq:stherm}, (black dashed line) or the three-parameters ansatz defined by Eqs.\eqref{eq:fonctions} and \eqref{eq:s} (red line).
    The dark grey zones mark regions where the kinetic energy $m(x/t)^2/2$ is greater than the transverse energy gap $\hbar \omega_\perp$. Atoms with such kinetic energies might populate transversely excited states. Since almost all the boundary profile lies between these grey zones, one expects that the physics is well captured by the  one-dimensional model. (b) Comparison of the occupation factors obtained for both fitted 
    occupation factor distributions. 
}
    \label{fig:fitted_border}
\end{figure}

The first ansatz that we try is the occupation factor of a Gibbs ensemble, where the fitting parameters are the temperature $T$ and the chemical potential $\mu$. This was calculated first by Yang and Yang~\cite{yang1969thermodynamics}, who showed that the occupation factor $\nu(\theta)$ is the solution of the integral equation 
\begin{equation}
\label{eq:fonctions}
     s'(\nu(\theta)) =  \frac{a}{b}
     -\frac{m\theta^2}{2 b} + \int d\theta' \Delta(\theta-\theta')
    \left [ s(\nu(\theta')) - \nu(\theta')s'(\nu(\theta)) \right ]
\end{equation}
where $a=\mu$, $b=k_B T$, the function $s:[0:1]\rightarrow {\mathbb R}$ is
\begin{equation}
    s(y)=  - y \ln y +(1-y)\ln(1-y) \, ,
  \label{eq:stherm}
\end{equation}
and $s'$ is its derivative. The integral $\int s(\nu(\theta))  
\rho_s(\theta) d\theta $ is the entropy per unit length of the occupation factor distribution $\nu(\theta)$~\cite{yang1969thermodynamics}.
For a given $T$ and $\mu$, Eq. \eqref{eq:fonctions} can be solved numerically iteratively very efficiently using the fact that 
$s'^{-1}(\epsilon)=1/(e^{\epsilon}+1)$.

In Fig.~\ref{fig:fitted_border} we compare the experimental boundary profile with the best fit obtained from this thermal equilibrium ansatz. For the data set shown here, the fitted temperature and chemical potential are 
$T=280 \pm 17 $nK and $\mu / k_{\rm B}=71.5 \pm 0.7 $nK. The ratio $k_B T /\mu$ is noticeably larger than that usually found in our experiment when fitting the density profile of harmonically confined clouds since in the latter case $k_B T /\mu_0$, where $\mu_0$ is the peak chemical potential takes values typically between 1 and 2~\cite{dubois_probing_2024,Fang2015}. We do not know where this discrepancy is coming from. It could originate from the fact that cooling in a quartic trap would be different from cooling in a harmonic trap.

The uncertainties in the fitting values of $T=280 \pm 17 $nK and $\mu / k_{\rm B}=71.5 \pm 0.7 $nK are the standard deviation error  obtained assuming that the discrepancy between the fit and the data is solely due to uncorrelated noise in the data points. However, discrepancy between the fit and the data shows some systematic effects, seen on all our datas, that indicate that the model does not faithfully describes the cloud: the left tail of the experimental data is wider than that of the fit while on the right side, the experimental date are more sharp. 
Interestingly, deviations from predictions for thermal states have already been observed for gases produced by evaporative cooling in an atom-chip setup~\cite{Fang2015,johnson_long-lived_2017}: the phonons are found to have a lower temperature than the high-energy excitations, the latter coinciding with the presence of rapities of large absolute value, 
as shown in~\cite{bouchoule_breakdown_2021}.

In order to obtain a better fit of the boundary profile, we relax the assumption that the initial could is at thermal equilibrium. 
Even if non-thermal, the cloud is however in a global stationarity stationnary state, since 
the density profile of the trapped cloud is time-independent.
General stationary states of the GHD equations in presence of a confining potential $V(x)$ have a local occupation 
factor distribution $\nu(x,\theta)$ which obeys Eq.~\eqref{eq:fonctions} with a global 
'temperature coefficient' $b$ and a local 'chemical potential coefficient'  $a(x)=a_0 -V(x)$, where $a_0$ is the central 'chemical potential'. However, general stationary states differ from thermal states by the choice of the 'entropy' function $s(y)$, which does not need to be given by  Eq.\eqref{eq:stherm}, but can be an arbitrary function~\cite{Bulchandani23}. In the following, we generalize the 
thermal stationary state by modifying the function $s$ as follows,
\begin{equation}
    s(y)=  -(1+cy)\left( y\ln(y) +(1-y)\ln(1-y) \right ) ,
    \label{eq:s}
\end{equation}
where $c$, dimensionless, is a third fitting parameter (together with the coefficients $a$ and $b$), which is vanishing 
for thermal states. The motivation for this ansatz is the following. First, inspecting Eq.\eqref{eq:fonctions}, one sees that modifying Eq.~\eqref{eq:stherm} using a global multiplicative factor $\alpha$ would keep a thermal state but with a temperature now equal to $\alpha b$ instead of $b$.
Very naively the above ansatz should tend to produce a "temperature" -- in a loose sense-- which depends on the occupation factor: the rapidities of large absolute value, whose occupation factor is small, would  be at a different "temperature" as the bulk of the rapidity distribution whose occupation factor is close to one, a feature seen in~\cite{johnson_long-lived_2017}. 

We used the ansatz Eq.\eqref{eq:s} to fit  our data.
 More precisely, we fit the experimental boundary profile 
with the prediction for 
an initial  occupation factor $\nu_0$ which is the one which obeys Eq.~\eqref{eq:fonctions} with the  function $s$ defined in Eq.\eqref{eq:s}.
The fit decreases the square distance
to the data by  $30$ \% compared to the thermal fit and it gives the optimal paramaters 
$a/k_B=74 \pm 0.2 $nK, $b/k_B=480 \pm 20$nK and 
$c=2.6 \pm 0.25 $. 

\section{Local rapidity distribution within the boundary}
\label{sec:local}

 For an initial state of the gas corresponding to a smooth occupation factor $\nu(\theta)$ ---for instance a thermal state---, the occupation factor  $\nu^*(x/t,\theta)$ at fixed ratio $x/t$ is expected to be highly asymmetric as a function of $\theta$, according to Eq.\eqref{eq:nuetoile} and as illustrated in Fig~\ref{fig:euler}. Indeed,  on the right side -- large $\theta$ -- it has a jump discontinuity, similar to the one of the ground state occupation function, while on the left side -- small $\theta$ -- it is smooth. 
 This feature of $\nu^*(x/t,\theta)$ induces a strong asymmetry of 
 the expected rapidity distribution $\rho(x/t,\theta)$.
 To reveal such peculiar features of the local state of the gas within the boundary, we use the protocol introduced in Ref.~\cite{dubois_probing_2024} to probe the local rapidity distribution. This protocol uses a technique to select a small slice of the gas. The rapidity distribution of the selected slice is then measured performing a 1D expansion~\cite{malvania_generalized_2021,wilson_observation_2020,yang2024phantom,PhysRevA.107.L061302,horvath_observing_2025}:
 after a sufficiently long 1D expansion one indeed expects that the density profile and the velocity distribution of the atoms become equal to the rapidity distribution~\cite{campbell_sudden_2015}.  To enhance the expected asymmetry of the local rapidity distribution, the data set used in this section corresponds to a cloud hotter than those used in the previous sections. The full experimental protocol is detailed below.
		
		First, we let the gas expand for a time $t=18 \mbox{ ms}$, such that the boundary broadens and covers a large zone of $\sim 350 \mu$m, see Fig.~\ref{fig:simul_deformation}(a).
		Then we select the slice of the gas that lies in an interval  $[x_0-\ell/2, x_0+\ell/2]$, removing all atoms lying outside the slice with a pushing beam~\cite{dubois_probing_2024}.
		In Fig.~\ref{fig:simul_deformation}(a) we show the density profile $1$ms after the selection of the slice.
		The fit to a smoothened rectangular  function gives $x_0= 18\,\mu$m.
		For calculations, the width $\ell$ will be determined using the 
		number of selected atoms (see below).
		Finally, we let this slice expand in 1D for an expansion time $\tau$, and then we measure the longitudinal density $\tilde{n}(x, {\tau} )$.  The latter reflects the total rapidity distribution of the slice $\Pi(\theta)=\int_{x_0-\ell/2}^{x_0+\ell/2}  \rho(x,\theta) dx$, because the asymptotic behavior as $\tau\rightarrow \infty$ is $ 
		\tau \tilde{n}(x, {\tau} ) \simeq \Pi((x-x_0)/\tau)$.  The 
		expected asymmetry of $\Pi$ is thus expected to induce an asymmetry of the density $\tilde{n}(x,\tau)$ as a function of the position $x$. We observe this asymmetry experimentally in our expansion profiles, as expected. This is shown in Fig. \ref{fig:simul_deformation}~(b) for an expansion time of $\tau=30~\mbox{ms}$.
		
		\begin{figure}[!htb]
		
		\end{figure} 
		
		\begin{figure}[!htb]
		
		\begin{tikzpicture}
            \node[rectangle, draw = none] (bord) at (0,0) {\includegraphics[width=0.5\textwidth]{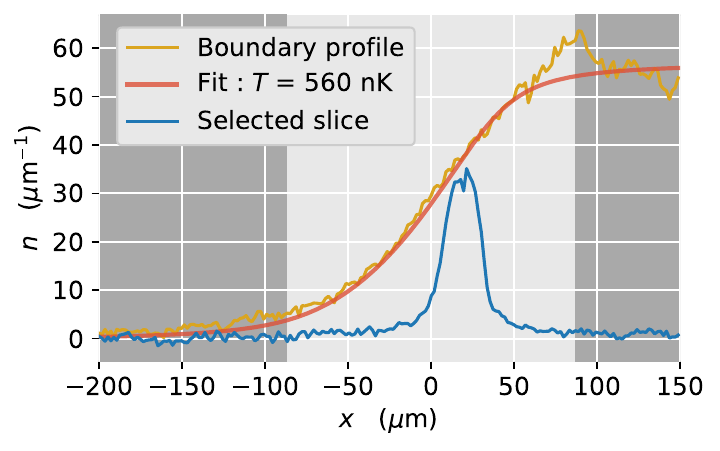}
            };
            \node[circle, draw=none, above=0cm of bord , shift={( -2.5cm , -0.5cm )} ] {(a)};
    
            \node[right=1mm of bord , shift={( -0.5cm , 0cm )}] (assy) {
              \includegraphics[width=0.5\textwidth]{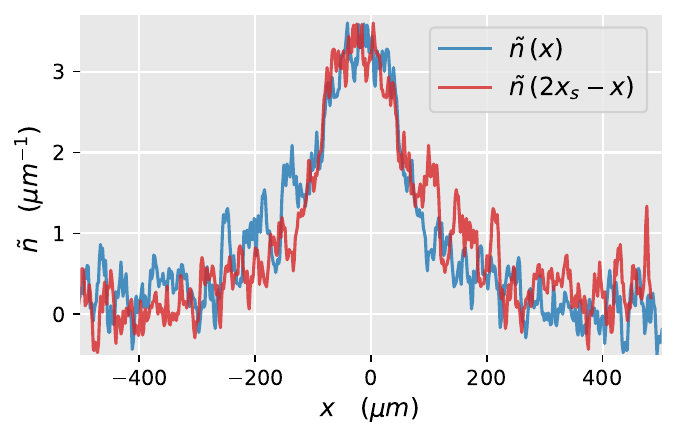}
            };
            \node[circle, draw=none, above=0cm of assy , shift={( -2.5cm , -0.5cm )}] {(b)};
        \end{tikzpicture}
        			\caption{(a). {\it Boundary profile and selected slice.} The boundary profile after  an evolution time $t=18\,ms$ is shown in solid yellow line. A thermal fit yielding $T=560\,nK$ is shown in orange.  The density profile taken $1\,ms$ after the slice selection is shown in blue. 
        			The dark grey area is the region where $m(x/t)^2/2 > \hbar\omega_\perp$. (b).  {\it Asymmetry of the slice expansion profile.} The density profile after an expansion of the slice aver a time $\tau=30$ ms is compared to its miror image. The symmetry center $x_s= -17\,\mu$m is the point that minimises the curves square distance $\delta^2=\int dx (\tilde{n}(x)-\tilde{n}(2x_s -x))^2$.  
        			}
        		\label{fig:simul_deformation}
        		
   			\end{figure}

		To go beyond this qualitative observation, we perform an Euler-scale GHD calculation
		of the expansion profile. 
		Since here we are not interested in the exact shape of the smooth borders of the initial occupation factor $\nu_0(\theta)$, we simply model it by a thermal distribution. A fit of the boundary profile before the selection of the slice,  shown in Fig. 6(a),  
        yields $T=560 \pm 22.5$ nK. Here we actually 
        perform a one-paramater fit: the chemical potential is not a fit parameter, instead, for each temperature, the 
        chemical potential is adjusted so that the linear density it corresponds to is  the linear density
        in the region $x>0$ measured before the boundary broadening.
		  Starting from the initial sharp profile, we simulate both the boundary broadening and the slice expansion with GHD, assuming a perfect slicing, i.e. $\nu(x,\theta)=0$ if $|x-x_0|>\ell/2$ and $\nu(x,\theta)$ is unchanged if $|x-x_0|<\ell/2$. The slice width $\ell$ is adjusted so that the calculated number of selected atoms equals  
		 the number of atoms in the experimental expansion profile, and we find $\ell= 24 \,\mu$m. 
		 
\begin{figure}[!htb]

		\begin{tikzpicture}
            \node[rectangle, draw = none] (exp) at (0,0) {
                \includegraphics[width=0.5\textwidth]{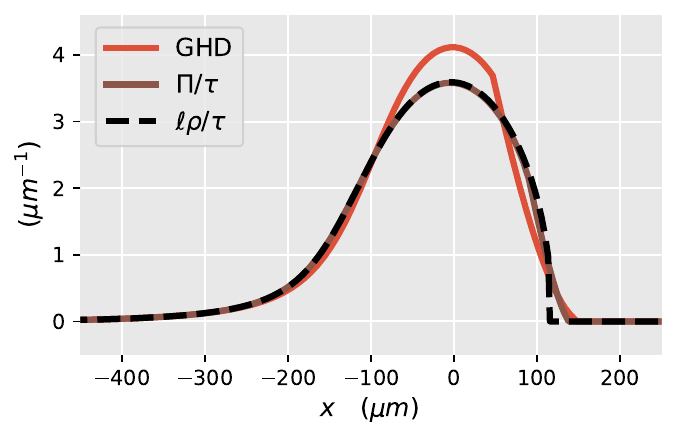}
            };
            \node[circle, draw=none, above=0cm of exp , shift={( -2.5cm , -0.5cm )} ] {(a)};
    
            \node[right=1mm of exp , shift={( -0.5cm , 0cm )}] (pi) {
                \includegraphics[width=0.5\textwidth]{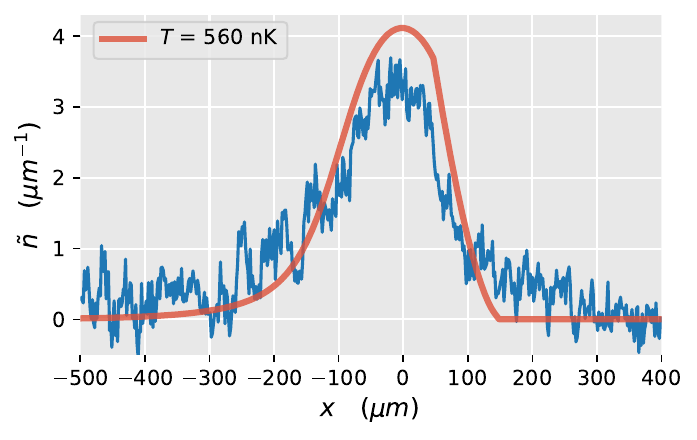}
            };
            \node[circle, draw=none, above=0cm of pi , shift={( -2.5cm , -0.5cm )}] {(b)};
            
        \end{tikzpicture}
    	
    		\caption{(a) {\it Density profile after slice expansion: effects of finite slice width and finite expansion time.} Orange line:  GHD calculation of the density profile after an expansion of the  slice for $\tau=30$~ms, using as the initial temperature the value $T=560$~nK obtained fitting the boundary profile.  The brown  line
    		is the expected distribution if one 
    		assumes that the asymptotic large expansion time is reached, {\it i.e.} it shows
    		 $\Pi((x-x_0)/\tau)/\tau$, where  $\Pi(\theta)=\int_{x_0-\ell/2}^{x_0+\ell/2} \rho(x,\theta) d x$ is the rapidity distribution of the selected slice. 
    		The  black dashed line is $\ell \rho(x_0, (x-x_0)/\tau)/\tau$, which is the expected result in the limit of large $\tau$ and for a slice width $\ell$ negligible compared to the boundary extension. 
    		(b) {\it Comparison to experimental data.} Experimental data obtained after slice expansion during $\tau=30$ ms (blue line) is  compared to GHD calculations assuming an initial thermal state.
    		 The  orange line, which is the same as in Fig.(a).
    		}
   			\label{fig:simul_expansion}					
		\end{figure}

		The simulated expansion profile is shown in Fig.~\ref{fig:simul_expansion}(a). It displays a strong asymmetry, as expected, with a sharp right edge and a vanishing density beyond a certain point on the right. The sharpness of this edge is, however, less pronounced than the one expected for the local rapidity distribution $\rho(x_0,\theta)$ at $x=x_0$, shown as dashed line in Fig.~\ref{fig:simul_expansion}(a). Two effects contribute to the broadening of  the  edge. First the rapidity distribution is not homogeneous inside the slice and $\Pi(\theta)$ differs from $\ell \rho(x_0,\theta)$, as seen comparing the 
		solid brown line and dashed line in Fig.\ref{fig:simul_expansion}~(a). Second, the expansion time is finite and the expansion profile is not exactly $\Pi((x-x_0)/\tau)/\tau$, as seen comparing the 
		red  and brown  solid lines in Fig.\ref{fig:simul_expansion}~(a).
		
		Next, we compare the expansion profile simulated with GHD to the experimental data.
		As shown in Fig~\ref{fig:simul_expansion}(b), the  predicted profile reproduces the main features of the  experimental expansion profile. Discrepancy are however as large as 25\% in the central part of the profile. 
		The most striking difference between the measured profile and the expected one is the presence of tails in the right edge of the experimental profile -- see the density profile in Fig.\ref{fig:simul_expansion}(b).
		Such tails are absent from the Euler-Scale GHD calculations because the occupation factor distribution inside the slice strictly vanishes above a certain rapidity. The reason for the presence of such tails is unclear. It might be due to edge effects associated to the slicing procedure, atoms at the edges of the slice being heated by the pushing beam. There is also maybe an effect of diffusion that go beyond Euler-scale GHD: the diffusive term, neglected within Euler-scale GHD, could have an impact  at the beginning of the edge deformation when gradients are large. Finally, we do not exclude that 3D effects are present since, for  the data analysed in this section, a non vanishing fraction of the atoms do have a longitudinal energy larger than $\hbar\omega_\perp$. This is seen in Fig.~\ref{fig:simul_deformation} which shows that the boundary profile is non vanishing in the region $\zeta > \sqrt{2\hbar\omega_\perp/m}$. One possible way to investigate this would be to adapt the method of the collision integral developed in Ref.~\cite{moller2021extension} to our setup (see also the discussion in Sec. 2.4.2 of the review~\cite{bouchoule_generalized_2022}).

\section{Conclusion}
We have investigated experimentally the bipartite quench protocol for a gas of bosons strongly confined transversely. We have checked that the time evolution
obeys the Euler hydrodynamic scaling since the density profile is found to be a function of $x/t$ (Fig.~\ref{fig:euler}). The density profile is close to the one predicted by the generalized hydrodynamic theory for the Lieb-Liniger gas at vanishing temperature,
the latter coinciding with the 
Gross-Pitaevski prediction for the parameters of our data.
The differences signal that the system is not in 
the ground state.
We showed that the measurement of the boundary profile $n(v)$ could in principle permit the reconstruction of the occupation factor $\nu(\theta)$ of the initial gas, realizing a generalized thermometry method. 
However, in practice, we prefer to fit the observed boundary profile with the one obtained from generalized hydrodynamics using an ansatz for the occupation factor.
We have found that the measured boundary profiles are not very well accounted for by a thermal 
occupation factor, and we have considered more general occupation factors corresponding to stationary trapped clouds, which give a better fit with our data. Finally, we present measurement of the local rapidity distribution inside the 
boundary. The data show the expected asymmetry of the distribution. The distribution however shows noticeable  differences compared to  the GHD predictions, whose 
origin is not elucidated.

This work calls for further  experimental investigations. In the near future we plan to compare the rapidity distributions
obtained with the  bipartite quench protocol by fitting the boundary profile
to the  rapidity distribution obtained using the 
slice expansion protocol~\cite{dubois_probing_2024}.
Moreover, the temperatures obtained in this paper by the thermal fits are large so that the
effects of populated transverse states might have an impact~\cite{moller2021extension,cataldini2022emergent}. Thus it would be interesting to investigate clouds at smaller energies. 
The investigation of the local rapidity distribution within the boundary
deserves further studies in order to elucidate the origin of the tails 
on the side that is expected to be effectively at vanishing entropy.

\section*{Acknowledgements}
We thank V.~Bulchandani for discussion about 
general stationary states of GHD in a confining potential and A.~Urilyon and J.~de Nardis 
for discussion about the possibility to observe 
diffusive effects. We also thank F. Nogrette, from LCF, for work on the installation of the DMD experiment and A.-L. Coutrot, from LCF, for reparation work on the chip. 

\paragraph{Funding information}
We thank Sophie Bouchoule, Alan
Durnez and Abdelmounaim Harouri of C2N laboratory
for the chip fabrication. C2N is a member of RENATECH, the French national network of large facilities for
micronanotechnology. This work was supported by ANR Project QUADY-ANR-20-CE30-0017-01.

\begin{appendix}

\section{Appendix: Attempt at reconstructing the occupation factor from the boundary profile}

In this appendix we elaborate on our attempt at reconstructing the occupation factor $\nu_0 (\theta)$ from the boundary $n(\zeta)$, briefly reported in Section~\ref{sec:retreving} in the main text.

\begin{figure}[ht]
    \centering
    \begin{tikzpicture}
        \draw (0,0) node{\includegraphics[width=0.4\textwidth]{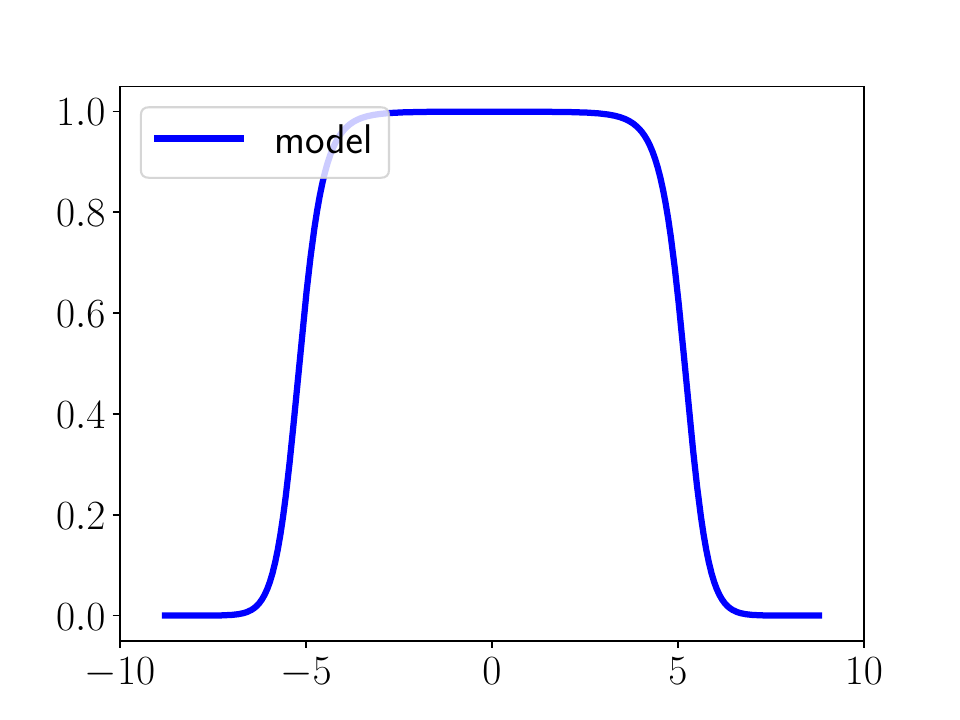}};
        \draw (-3,0) node[rotate=90]{\small occupation factor $\nu_0 (\theta)$};
            \draw (0,-2.3) node {\small rapidity $\theta$ ($\mu$m/ms)};
            \draw (8,0) node{\includegraphics[width=0.4\textwidth]{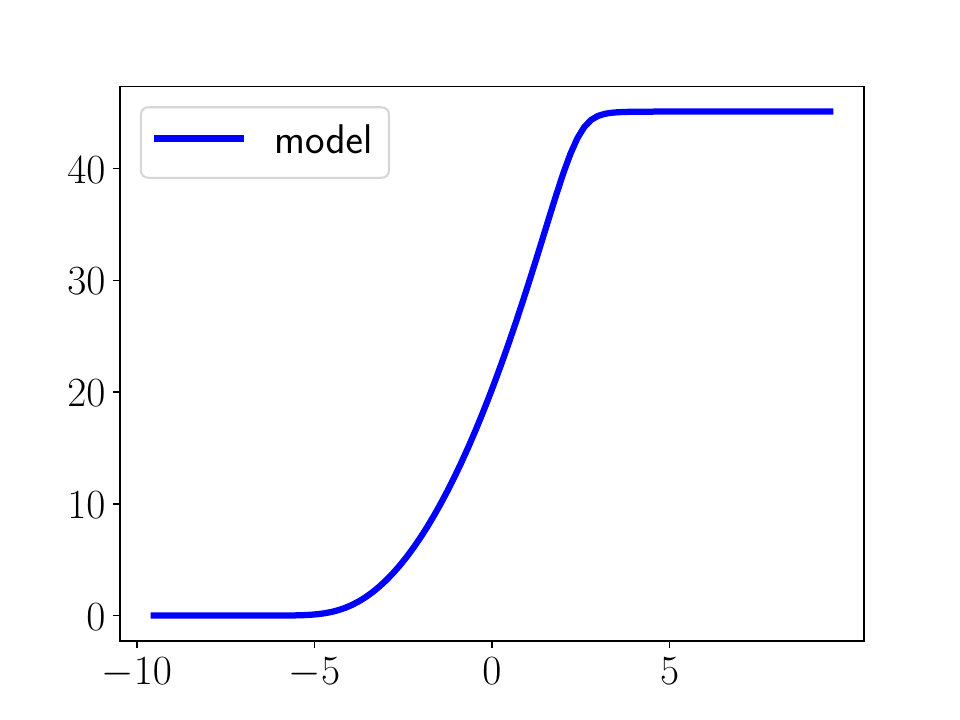}};
            \draw (5,0) node[rotate=90]{\small atom density $n(\zeta)$  ($\mu$m$^{-1}$)};
            \draw (8,-2.3) node {\small $\zeta = x/t$ ($\mu$m/ms)};
            \draw[->,thick] (3,0) -- ++(1.2,0);
        \draw (4,-7) node{\includegraphics[width=0.5\textwidth]{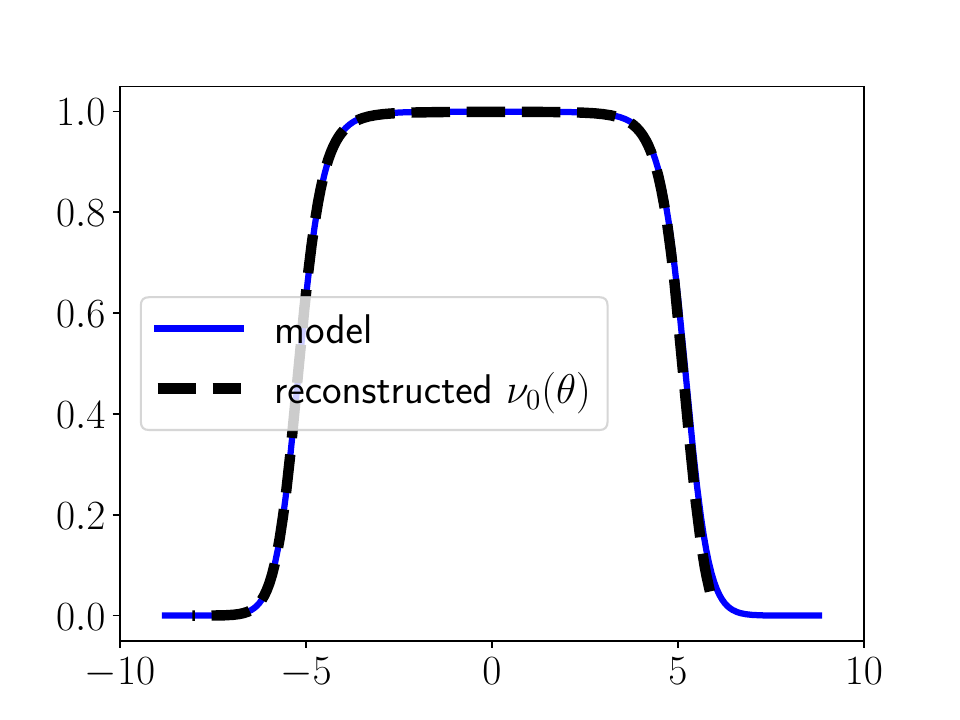}};
        \draw[->,thick] (8,-2.8) -- (4,-4.5);
        \draw (4,-3.2) node {reconstruction of $\nu_0(\theta)$};
        \draw (4,-3.7) node {with algorithm};
        \draw (0.3,-7) node[rotate=90]{\small occupation factor $\nu_0 (\theta)$};
            \draw (4,-9.9) node {\small rapidity $\theta$ ($\mu$m/ms)};
    \end{tikzpicture}
    \caption{Test of the algorithm described in Section~\ref{sec:retreving} on a model occupation ratio of the form $1/(1+e^{a \theta^2 - b})$. Upper left: the model occupation ratio [Parameters: $a=0.3 {\rm ms}^2 . \mu {\rm m}^{-2}$ and $b=8$.] Upper right: the corresponding boundary profile $n(\zeta)$. Bottom: the reconstructed occupation ratio obtained with the algorithm, compared with the model occupation factor. [The coupling constant $g$ and atom mass $m$ are the same as in the main text.]}
    \label{fig:app_reconstruction_check}
\end{figure}

First, we test the algorithm detailed in Section~\ref{sec:retreving}. For this we take a model occupation function of the form $1/(1+ e^{ a \theta^2 - b })$, and compute the corresponding `model boundary profile' $n(\zeta)$. Then we run our algorithm on this model boundary profile to reconstruct the occupation factor $\nu_0(\theta)$. We see in Fig.~\ref{fig:app_reconstruction_check} that the reconstruction works as it should.

\begin{figure}[ht]
    \centering
    \begin{tikzpicture}
        \draw (0,0) node { \includegraphics[width=0.47\textwidth]{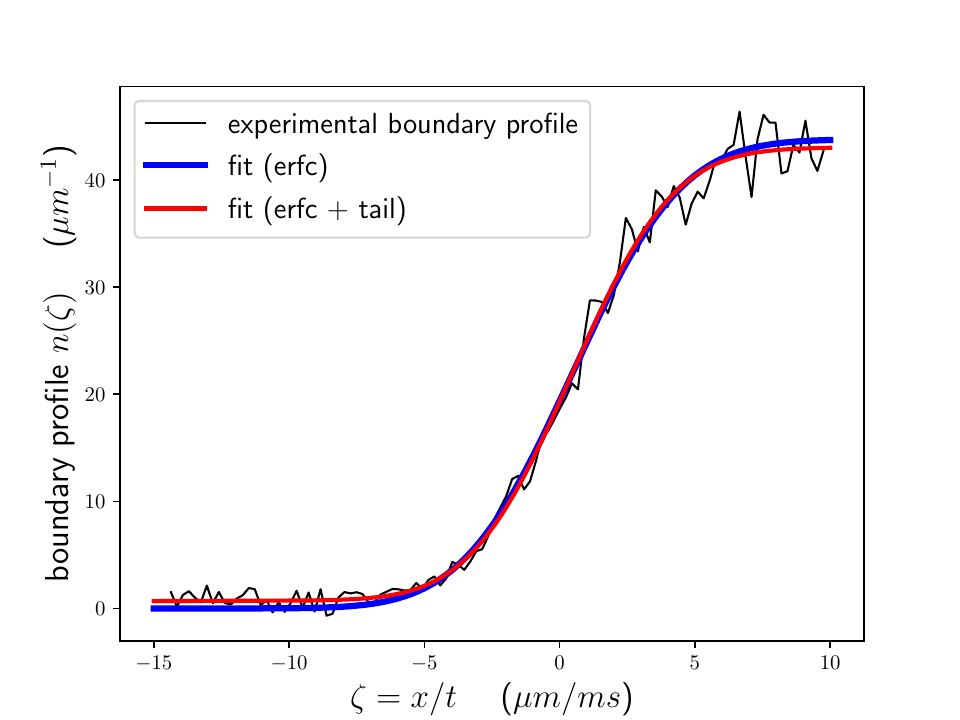}};
        \draw (8.4,0) node {
    \includegraphics[width=0.47\textwidth]{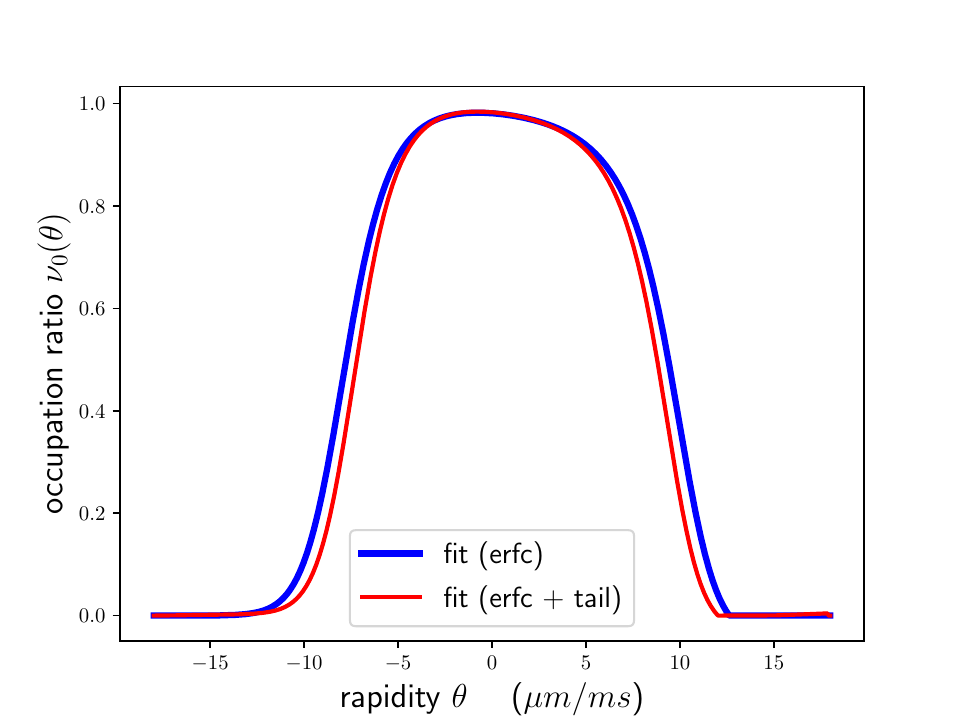}};
        \draw[->, thick] (3.5,0) -- (4.5,0);
        \draw (4,0.4) node{{\footnotesize algorithm}};
    \end{tikzpicture}
    \caption{Left: fit of the experimental data for the density profile (noisy black line) with an error function $a \, {\rm erfc}(-b*(\zeta-c))$ (blue line) and with an error function plus a slowly decaying tail on the left, of the form $a  \, {\rm erfc}(-b*(\zeta-c)) -  d/(\zeta-e)$. Right: the corresponding occupation factors reconstructed with the algorithm of Section~\ref{sec:retreving}. We conclude that the method is not satisfactory for our purposes: the reconstructed occupation ratio is in general not symmetric and is too sensitive to the details of the chosen ansatz for the fit of the density profile.}
    \label{fig:fit_erfc}
\end{figure}

Next, we try to apply the algorithm to our experimental boundary profiles. We immediately face the following problem: The theoretical prediction for the  boundary profile $n(\zeta)$ is necessarily monotonously increasing with $\zeta$, while the experimental data are noisy so they can locally decrease. Therefore, in order to get a monotonously increasing function from our data, we start by fitting the data with a simple ansatz of the form $a \, {\rm erfc}(-b*(\zeta-c))$, where ${\rm erfc}$ is the error function, see Fig.~\ref{fig:fit_erfc} (left panel). We have also tried to modify the ansatz in order to allow for a slower decay of the tail for $\zeta<0$, see Fig.~\ref{fig:fit_erfc}. However the signal-to-noise ratio in our data does not allow us to discriminate between different choices of fit functions.

Once we have fitted the boundary profile $n(\zeta)$, we can run the algorithm on $n(\zeta)$ to find the corresponding occupation factor $\nu_0(\theta)$, see Fig.~\ref{fig:fit_erfc} (right). Unfortunately we find that this method is not satisfactory. The occupation ratio should be symmetric under $\theta \rightarrow -\theta$, but we can see in Fig.~\ref{fig:fit_erfc} (right) that it is not, with our choice of fit functions for $n(\zeta)$. Moreover we observe that the reconstructed occupation ratio is quite sensititive to the details of $n(\zeta)$.We have concluded that our data are too noisy to exploit this method, and this is the reason why we turn to the alternative method explained in the main text.

\end{appendix}

\bibliography{Biblio_Lea.bib,Domain_Wall_paper}

\nolinenumbers

\end{document}